\documentclass{article}
\usepackage{enumerate,graphicx,natbib,amsfonts,amsbsy}
\usepackage[mathscr]{eucal}
\DeclareMathSymbol{\nrightarrow}{\mathrel}{AMSb}{"39}
\newcommand{\cequals}{\ensuremath{:\!\mbox{$=$}}\,}
\newcommand{\fI}{\mathfrak{I}}
\newcommand{\bx}{\boldsymbol{x}}
\newcommand{\eusF}{\mathscr{F}}
\newcommand{\curm}{{\sf m}}
\newcommand{\mcB}{{\mathcal{B}}}
\newcommand{\mcL}{{\mathcal{L}}}
\newcommand{\tA}{{{\mbox{\tiny A}}}}
\newcommand{\tB}{{{\mbox{\tiny B}}}}
\newcommand{\tG}{{{\mbox{\tiny G}}}}
\newcommand{\tE}{{{\mbox{\tiny E}}}}
\newcommand{\tD}{{{\mbox{\tiny D}}}}
\newcommand{\tT}{{{\mbox{\tiny T}}}}
\newcommand{\mx}{{{\mbox{\tiny Max}}}}
\newcommand{\tmN}{{{\mbox{\tiny$N$}}}}
\newcommand{\prb}{{\sf Prob}}
\newcommand{\prbs}{{\sf p}}
\newcommand{\ttime}{{\sf T}}
\addtocounter{footnote}{1}
\title{Boltzmann, Gibbs and the Concept of Equilibrium}
\author{ David A.\ Lavis\footnote{Department of Mathematics, King's College,
London WC2R 2LS, U.K. Email: DAVID.LAVIS@KCL.AC.UK}
\footnote{This modified version of the paper presented at the 20th Biennial Meeting of the
Philosophy of Science Association, Vancouver, Canada, November 2006 will appear in
{\em Philosophy of Science}. I am grateful to
the organizers of the conference for giving me the opportunity to contribute to the conference
and to Roman Frigg for many useful discussions.}}
\date{}
\begin{document}
\bibliographystyle{dcu}
\maketitle
\begin{abstract}
The Boltzmann and Gibbs approaches to statistical mechanics
have very different definitions of equilibrium and entropy.
The problems associated with this
are discussed and it is suggested that they
can be resolved, to produce a version of statistical mechanics
incorporating both approaches, by redefining equilibrium not
as a binary property (being/not being in equilibrium) but
as a continuous property (degrees of equilibrium) measured by the
Boltzmann entropy and by introducing the idea of thermodynamic-like
behaviour for the Boltzmann entropy. The Kac ring model is
used as an example to test the proposals.
\end{abstract}
\section{Introduction}\label{intro}
 The object of study in the Gibbs formulation of statistical mechanics is an ensemble of systems
and the Gibbs entropy is a functional of the ensemble probability density function.
 Equilibrium is defined as the state where the probability density function is a
  time-independent solution of Liouville's equation. The development of this
   approach has been very successful, but its extension to non-equilibrium
   presents contentious problems.\footnote{The most developed programme for
   doing this is that of the Brussels--Austin School of the late Ilya Prigogine.
  \citep[For a comprehensive review see][]{bish3}.}

To implement the Boltzmann approach \citep{leb1,bric,gold1} the phase space is
divided into a set of macrostates. The Boltzmann entropy at a particular point
in phase space is a measure of the volume of the macrostate in which the phase
 point is situated. The system is understood to be in equilibrium when the phase point is in
a particular region of phase space. The entropy and equilibrium are thus properties of a single system.

The purpose of this paper is to attempt to produce a synthesis of the
 Gibbs and Boltzmann approaches, which validates the Gibbs approach,
 as currently used in `equilibrium' statistical mechanics and solid
  state physics, while
at the same time endorsing the Boltzmann picture of the time-evolution of entropy,
 including `the approach to equilibrium'.  In order to do this we need to
resolve in some way three questions, to which the current versions
of the Gibbs and Boltzmann approaches offer apparently
irreconcilable answers: (a) What is meant by equilibrium? (b) What is statistical mechanical entropy?
and (c) What is the object of study?
The attempt to produce conciliatory answers  to (a) and (b) will occupy most of this paper. However,
we shall at the outset deal with (c). As indicated above, ensembles are an intrinsic feature of the
 Gibbs approach \citep[see, for example][8]{prig1}. However we  follow
the Neo-Boltzmannian view of \citet[38]{leb1}  that we ``neither have nor do we need ensembles $\ldots$''.
The object of study in statistical mechanics
is a {\em single system} and all talk of ensembles can be understood as just a way of giving
a relative frequency flavour to the probabilities of events occurring in that system.

We now describe briefly the dynamics and thermodynamics of the system together
with the statistical approach of Gibbs. The Boltzmann approach is described in greater
detail in Sect.\ \ref{pwtba}.

\paragraph{At the microscopic (dynamic) level}\hspace{-0.3cm} the system (taken to consist of $N$ microsystems)
is supposed to have a one-to-one autonomous dynamics $\bx\rightarrow \phi_t\,\bx$, on its phase space $\Gamma_{\tmN}$.
The system is reversible; meaning that there exists a self-inverse operator $\fI$ on
 $\bx\in\Gamma_{\tmN}$, such that $\phi_t\bx=\bx'$ $\rightarrow$ $\phi_t \fI\,\bx'=\fI\,\bx$.
  Then $\phi_{-t}=(\phi_t)^{-1}=\fI\phi_t\fI$.
On the subsets of $\Gamma_{\tmN}$ there is a sigma-additive measure $\curm$,
such that (a) $\curm(\Gamma_{\tmN})$ is finite, (b) $\curm$ is absolutely continuous with respect to
 the Lebesque measure on $\Gamma_{\tmN}$, and (c) $\curm$ is preserved by $\{\phi_t\}$; that is
$\curm(\phi_t\gamma)=\curm(\gamma)$, $\forall$ $t$ and measurable $\gamma\subset\Gamma_{\tmN}$.
This means that there will be no convergence  to an attractor (which could in the dynamic sense be taken as an
equilibrium state).

\paragraph{At the phenomenological (thermodynamic) level}\hspace{-0.3cm} equilibrium is a state
 in which there is no perceptible change in
macroscopic properties. It is such that a system: (a)
either is or is not in equilibrium (a binary property), (b)
never evolves out of equilibrium and (c)
when not in equilibrium evolves towards it.

\paragraph{At the statistical level (in the Gibbs approach)}\hspace{-0.3cm} the phase-point
$\bx\in\Gamma_{\tmN}$ is distributed according to a probability density function $\rho(\bx;t)$
invariant under $\{\phi_t\}$; meaning that it is a solution of Liouville's equation.
At equilibrium the Gibbs entropy is the functional
\begin{equation}
S_{\tG}[\rho]\cequals-k_{\tB}\int_{\Gamma_{\tmN}}\rho(\bx)\ln[\rho(\bx)]\,\mathrm{d}\curm
\label{giben}
\end{equation}
of a time-independent probability density function.
Problems arise when an attempt is made to extend the use of (\ref{giben}) to
non-equilibrium situations, which are now perceived as being
those where $\rho$ is time-dependent.
\section{The Macroscopic Level -- Boltzmann Approach}\label{pwtba}
Here we must introduce a set $\Xi$ of
macroscopic variables at the observational level
which give more detail than the thermodynamic  variables,
and a set of macrostates $\{\mu\}$ defined so that:
(i) every $\bx\in\Gamma_{\tmN}$ is in exactly one macrostate denoted by $\mu_{\bx}$,
(ii) each macrostate corresponds to a unique set of values for $\Xi$, (iii)
$\mu_{\bx}$ is invariant under all permutations of the microsystems, and
(iv) the phase points $\bx$ and $\fI\bx$ are in macrostates of the same
size.\footnote{When $\Gamma_\tmN$ is the direct
product of the configuration space and the momentum space, and
the macrostates are generated from a partition of the one-particle configuration space,
the points $\bx$ and $\fI\bx$ are in same macrostate. However, for discrete-time
systems phase-space is just configuration space\label{disc} and
the  points $\bx$ and $\fI\bx$ are usually in different macrostates.}
The Boltzmann entropy, which is a function on the macrostates, and consequently also
 a function on the phase points in $\Gamma_{\tmN}$, is
\begin{equation}
S_{\tB}(\bx)= S_{\tB}(\mu_{\bx})\cequals k_{\tB}\ln[\curm(\mu_{\bx})].\label{bolen}
\end{equation}
This is, of course, an extensive variable and the quantity of interest
is the dimensionless entropy per microsystem $s_{\tB}\cequals S_{\tB}/(Nk_{\tB})$, which
for the sake of brevity we shall refer to as the Boltzmann entropy.
Along a trajectory $s_{\tB}$ will not be a monotonically increasing function
of time. Rather we should like it to exhibit {\em thermodynamic-like} behaviour,
defined in an informal (preliminary) way as follows:
\begin{quote}
Definition (TL1):\textsl{The evolution of the system will be \textbf{thermodynamic-like}
if $s_{\tB}$ spends most of
 the time close to its maximum value,
from which it exhibits frequent small fluctuations and rarer large fluctuations.}
\end{quote}
This leads to two problems which we discuss in Sects.\ \ref{hdwde} and \ref{wntlt}.
\subsection{How Do We Define Equilibrium?}\label{hdwde}
Is it possible to designate a part of $\Gamma_{\tmN}$ as the equilibrium state? On the grounds
of the system's reversibility and recurrence we can, of course, discount the possibility that
such a region is one which, once entered by the evolving phase point,
will not be exited. As was well-understood by both Maxwell and Boltzmann, equilibrium must
be a state which admits the possibility of fluctuations out of equilibrium.
\citet[34]{leb1} and \citet[8]{gold1} refer to a particular macrostate as the equilibrium
macrostate and the remark by \citet[179]{bric},\label{briclab} that ``by far the
largest volumes [of phase space] correspond to the {\em equilibrium values} of the
macroscopic variables (and this is how `equilibrium' should be
defined)'' is in a similar vein. So is there a single equilibrium macrostate? If so it must be
that in which the phase point spends more time than in any other macrostate and, if the system were ergodic,
it would be the largest macrostate $\mu_{\mx}$ (see Sect.\ \ref{wntlt}),
with largest Boltzmann entropy. There is one immediate problem associated with this. Suppose we consider the
set of entropy levels $\{s_{\tB}(\mu)\}$, $\forall$ $\mu\subset\Gamma_{\tmN}$. Then, as has been shown by \citet{lavis5}
for the baker's gas, associated with these levels there may be degeneracies $\omega(\mu)$, such that, for some
$\mu$ with $\curm(\mu)<\curm(\mu_{\mx})$, $\curm(\mu)\omega(\mu)>\curm(\mu_{\mx})$. The effect
of this is that the entropy will be likely, in the course of evolution, to spend more time in a level less than
the maximum  \citep[see][Figs.\ 4 and 5]{lavis5}.

\begin{figure}[t]
\begin{center}
\includegraphics[width=120mm,angle=0]{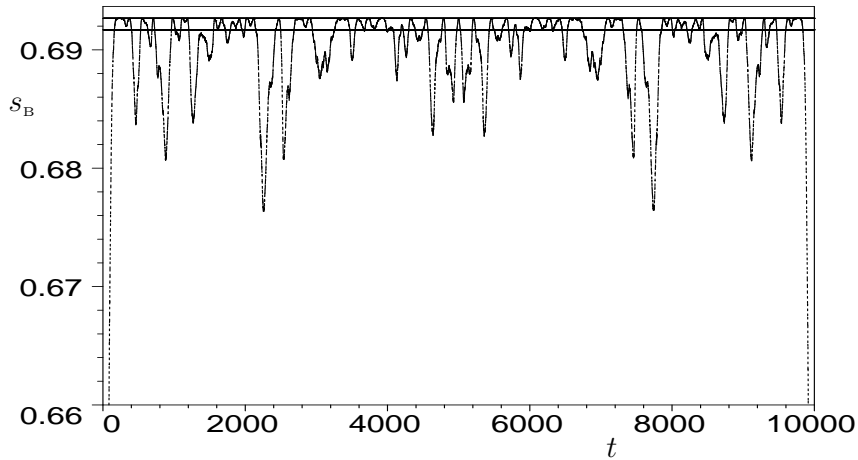}
\end{center}
\caption{Half a cycle of the plot of $s_{\tB}$
against time, for a Kac ring of $10,000$ spins. There are $m=509$ spin-flippers, distributed
randomly around the ring and the evolution is initiated in the low-entropy state
$k=4993$. The upper and lower horizontal lines correspond respectively to
 $s_\tB(\mu_{\mx})=0.69266408$ and $s_\tB(\mu_{221})=0.69168704$.}\label{fig1}
\end{figure}

Another example, which has been used to discuss the evolution
of Boltzmann's entropy \citep[see][Appendix 1]{bric} and which we shall use as an illustrative example in this paper,
is the Kac ring model \citep[][99]{kac2}.\footnote{This model
 consists of $N$ up or down spins distributed equidistantly around
  a circle. Randomly distributed at some of the midpoints between the spins are $m$ spin flippers.
The dynamics consists of rotating the spins (but not the spin-flippers) one spin-site in the clockwise direction,
with spins changing their direction when they pass through a spin flipper.
$\Gamma_{\tmN}$ consists of the $2^\tmN$ points, corresponding to all combinations of the two spin states,
and is decomposable into dynamically invariant cycles. If $m$ is even the parity
of $k$ is preserved along a cycle which has maximum size $N$. If $m$ is odd the parity of $k$ alternates with
steps along a cycle and the maximum cycle size is $2N$. }
Macrostates in this model can be indexed $k=0,1,\ldots,\frac{1}{2}N$, where $\mu_k$ is
the macrostate with $\frac{1}{2}N+k$ up spins and $\frac{1}{2}N-k$ down spins, giving
\begin{equation}
\curm(\mu_k)=  \frac{N!}{\left(\frac{N}{2}+k\right)!
\left(\frac{N}{2}-k\right)!}\hspace{0.7cm}\mbox{   with   }\hspace{0.7cm} \curm(\Gamma_\tmN)=2^{\tmN}.
\label{kac1}
\end{equation}
Then, of course, $\mu_\mx=\mu_0$ with $\curm(\mu_k)$ monotonically
 decreasing with increasing $|k|$. But, although the maximum macrostate
 is unique, $\omega(\mu_k)=2$, $\forall$ $k\ne 0$ and the entropy level corresponding to
the largest volume of $\Gamma_{\tmN}$ is given, for $N>2$, by the macrostate pair $\mu_{\pm 1}$. It may be suppose that
this question of degeneracy is an artifact of relevance only for small $N$. It is certainly the case, both
for the baker's gas and Kac ring, that, if $\mu'$ is a macrostate which maximizes $\curm(\mu')\omega(\mu')$,
although $\curm(\mu')\omega(\mu')\nrightarrow\curm(\mu_{\mx})$,
$s_{\tB}(\mu')\rightarrow s_{\tB}(\mu_{\mx})$,  as $N\rightarrow\infty$.  So, maybe the union of $\mu_\mx$ and
all the equally-sized macrostates with measure $\curm(\mu')$ can be used as the equilibrium state. To test this
possibility take the Kac ring and consider the partial sum
\begin{equation}
\curm(\mcB(N,k))\cequals\sum_{j=-k}^{k} \curm(\mu_j)= \curm(\mu_0)+2\sum_{j=1}^{k} \curm(\mu_j).\label{kac2}
\end{equation}
where $\mcB(N,k)$ is the union of all $\mu_j$ with $j\in[-k,k]$.
Then it is not difficult to show that, for fixed $k$,
\begin{equation}
\frac{\curm(\mcB(N,k))}{\curm(\Gamma_\tmN)}\rightarrow 0, \hspace{1cm}\mbox{as $N\rightarrow\infty$.}
\label{kac1bis}
\end{equation}
The proportion of $\Gamma_{\tmN}$ contained within the band of macrostates $\mcB(N,k)$
decreases with $N$.\footnote{For later reference we note that, in particular, this result
applies to $k=0$ with $\curm(\mu_0)/\curm(\Gamma_\tmN)\simeq\sqrt{2/N\pi}$ for large $N$.
 The proportion of phase space in the largest macrostate {\em decreases} with
$N$.} If we want the equilibrium region to satisfy the property described in the quote given above from
\citet{bric}, then this cannot be done by designating a band of a fixed number of macrostates in this way.
To ensure that as $N$ becomes large ``by far the
largest volumes [of phase space] correspond to the {\em equilibrium values} of the
macroscopic variables'' we need to choose a value of $k$ increasing with $N$. Thus for example
to create an `equilibrium band of macrostates' containing $99.999\%$ of $\Gamma_{\tmN}$ we must choose
$k=22$, for $N=100$, $k=70$, for $N=1000$ and $k=221$ for $N=10,000$. An entropy profile for this last
case is shown in Fig.\ \ref{fig1}.
The putative equilibrium state, representing $99.999\%$ of $\Gamma_{10,000}$ is
 given by the region bounded by the
horizontal lines in the figure. If the system were ergodic (like the baker's gas)
we would expect the system to spend almost all
of the 10,000 time steps in this region, whereas in this particular simulation only 4658 steps
satisfied this condition. This is to be expected as the Kac ring
is not ergodic, but has an ergodic decomposition into cycles of which
this figure represents half a cycle.\footnote{This is because $m$ is odd.
The second half of the cycle in which the spins are reverse has an identical entropy
profile.} The proportion of `equilibrium states' will differ between cycles.
So the problems with defining an equilibrium region are:
\begin{enumerate}[(i)]
\item Just choosing the largest macrostate as the equilibrium region, does not guarantee that
this region becomes an increasing proportion of phase space as $N$ increases. In fact the reverse
is the case for the Kac ring.
\item Any choice of a collection of macrostates to represent equilibrium is:
\begin{enumerate}[(a)]
\item Arbitrary: leading to an arbitrary division between fluctuations {\em within}
and {\em out of} equilibrium, as is shown in the profile in Fig.\ \ref{fig1}.
\item Difficult: as we have shown in the Kac ring. Except for ergodic systems there is no
clear way to make a choice which guarantees that the system will spend most of its time in equilibrium.
The choice of a region consisting of $99.999\%$ of phase space still yields an evolution
where only about $47\%$ of the points on the trajectory lie within it.
\end{enumerate}
\end{enumerate}
But why define equilibrium in this binary way? We suggest that
the quality which we are trying to capture is a matter of degree,
rather than the two-valued property of either being {\em in equilibrium} or
{\em not in equilibrium}.
We, therefore, make the following proposal:
\begin{quote}
Definition (C):\textsl{All references to a system being, or not being,
in equilibrium should be replaced by references to
the \textbf{commonness} of the state of the system, with this property being measured by (some possibly-scaled
form of) the Boltzmann entropy.}
\end{quote}
\subsection{We Need Thermodynamic-Like Behaviour to be\break Typical}\label{wntlt}
By this we mean that most initial states of the system should lead to\break thermodynamic-like behaviour.
Before discussing the dynamic properties needed for this, we shall
refer briefly to the more limited notion of typicality employed by the Neo-Boltzmannians
and contained
in the assertion \citep[S348]{leb3}  ``that $S_{\tB}$ will {\em typically} be increasing in a way which
{\em explains} and describes qualitatively the evolution towards equilibrium of macroscopic systems''.
The conditions for this to be the case were first discussed by \citet[32--34]{ehr2}.

Consider a macrostate $\mu$ divided
into four parts $\mu^{(--)}$, $\mu^{(-+)}$, $\mu^{(+-)}$, $\mu^{(++)}$, where $\mu^{(-+)}$ consists
of those points in $\mu$ which have evolved from a smaller macrostate and which evolve into a larger
macrostate, with the other parts defined in a similar way.\footnote{For simplicity
 we have excluded evolutions between macrostates of the same size.}
On grounds of symmetry, if $\bx$ and $\fI\bx$ are both in $\mu$, then $\curm(\mu^{(-+)})=\curm(\mu^{(+-)})$;
otherwise the macrostates will be in reversal pairs with plus and minus signs interchanged. In order
for forward evolution to a larger macrostate to be typical it must be the case that the overwhelmingly
largest part of $\mu$ is $\mu^{(++)}$. This was asserted without proof by  \citet[33]{ehr2}.
More recent arguments have been advanced from the point of view that a macrostate is more likely to be surrounded by
larger macrostates or that it is easier to `aim at' a larger, rather than a smaller, neighbouring macrostate. Even accepting this argument,
it can at the most explain how, if the state of a system is assigned randomly to a point in a macrostate,
then the subsequent first transition is to a larger macrostate. As was pointed out by \citet{lavis5},
it gives no explanation for the entropy direction at the next transition, since the part of the macrostate
occupied after the first transition will be determined by the dynamics. In any case, we wish to
argue that this is a too narrowly defined version of typicality, which should be applied to thermodynamic-like
behaviour over the whole evolution.

We have already argued in Sect.\ \ref{hdwde} that commonness or `equilibriumness' is a matter
of degree and it is clear that thermodynamic-like behaviour is also a matter of degree,
for which we need to proposed a measure. A difference
between these properties is that commoness is something which can be assessed at an instant
of time, whereas thermodynamic-like behaviour is a temporally
global property assessed over the whole trajectory.

Let $\mcL_{\bx}$ be a trajectory in $\Gamma_{\tmN}$ identified (uniquely) by the property that
it passes through the point $\bx$ and let $\ttime_{\bx}(\gamma)$ be the proportion of time
which the phase point evolving along $\mcL_{\bx}$ spends in the some
$\gamma\subset\Gamma_\tmN$.\footnote{It was shown by \citet{birk} that
$\ttime_{\bx}(\gamma)$ exists and is independent of the location of $\bx$ on $\mcL_{\bx}$
for almost all (f.a.a.) $\bx\in\Gamma_\tmN$; that is, except possibly
for a set of $\curm$-measure zero. From this it follows \citep[see e.g.][]{lavis1}
that $\ttime_{\bx}(\gamma)$ is a constant of motion f.a.a. $\bx\in\Gamma_\tmN$.}
Definition TL1 is an informal qualitative
definition of thermodynamic-like behaviour, for which we need the entropy
profile of $s_{\tB}$, not only to be quite close to
$(s_{\tB})_\mx\cequals s_{\tB}(\mu_\mx)$ for most of its evolution, but also for fluctuations around this value
to be fairly small. We, therefore, propose the following definition:
\begin{quote}
Definition (TL2):\textsl{The degree to which the evolution of the system is \textbf{thermodynamic-like}
along $\mcL_{\bx}$ is measured by the extent to which
\begin{eqnarray}
\triangle_{\bx}[s_{\tB}]&\cequals&|\langle s_{\tB}\rangle_{\bx}-(s_{\tB})_{\mx}|,\label{wntlt1}\\
&&\hspace{-4.8cm}\mbox{and}\nonumber\\
\Psi_\tmN[s_{\tB}]&\cequals
&\sqrt{\langle\left[s_{\tB}-\langle s_{\tB}\rangle_{\bx}\right]^2\rangle_{\bx}},\label{wntlt2}
\end{eqnarray}
are small, where
\begin{equation}
\langle s_{\tB}\rangle_{\bx}\cequals\sum_{\{\mu\}}\ttime_{\bx}(\mu)s_{\tB}(\mu),\label{wntlt3}
\end{equation}
is the time-average of $s_{\tB}$ along  $\mcL_{\bx}$ and $\Psi_{\bx}[s_{\tB}]$ is the
standard deviation with respect to the time distribution.}
\end{quote}
Of course, it could be regarded as unsatisfactory that {\em two} parameters are used as a measure
of the degree of a property and it is a matter of judgement which is more important.
For  the Kac ring of 10,000 spins with the entropy profile shown in Fig.\ \ref{fig1},
\begin{equation}
\triangle_{\bx}[s_{\tB}]= 0.58122724\times 10^{-2},\hspace{0.7cm}
 \Psi_{\bx}[s_{\tB}]=0.31802804\times10^{-1}
\label{wntlt5}
\end{equation}
and, as a comparison, for the same ring with the flippers placed
at every tenth site
\begin{equation}
\triangle_{\bx}[s_{\tB}]=0.20078055,\hspace{0.7cm} \Psi_{\bx}[s_{\tB}]=0.20632198.
\label{wntlt6}
\end{equation}
It is clear (and unsurprising) that the random distribution of spin flippers
leads to more thermodynamic-like behaviour.

To explore the consequences of TL2 we distinguish between four aspects of a system:
\begin{enumerate}[(i)]
\item The number of microsystems $N$ and their degrees of freedom, together giving
the phase space $\Gamma_\tmN$, with points representing microstates.
\item The measure $\curm$ on $\Gamma_\tmN$.
\item The mode of division of $\Gamma_\tmN$ into the set $\{\mu\}$ of macrostates.
\item The $\curm$-measure preserving dynamics of the system.
\end{enumerate}
Having chosen (i) and (ii) the choices for (iii) and then (iv) are not unique.
In the case of the baker's gas \citep{lavis5}, $\Gamma_\tmN$
is a $2N$--dimensional unit hypercube with volume measure. Macrostates are specified by partitioning
each square face of the hypercube. With such a setup it would now be possible to choose
all manner of discrete-time dynamics.\footnote{Discrete time because
of the absence of a momentum component in the phase space. (See the footnote on page \pageref{disc}.)}

Whether a system is ergodic will be determined by (i), (ii) and (iv) and, if it is,
\begin{equation}
\ttime_{\bx}(\gamma)= \tau(\gamma), \hspace{0.7cm}\mbox{where}\hspace{0.7cm}\tau(\gamma)
\cequals\frac{\curm(\gamma)}{\curm(\Gamma_{\tmN})},\hspace{1cm}\mbox{$\forall$\hspace{0.1cm} $\gamma\subset\Gamma_\tmN$,}
\label{ttb1}
\end{equation}
is both the proportion of $\Gamma_\tmN$ in $\gamma$ and of the time spent in $\gamma$,
f.a.a. $\mcL_{\bx}\in\Gamma_\tmN$. This will be the case for the baker's gas but not the Kac ring.
For any specification of (i)--(iii) we denote the results of computing
$\langle s_{\tB}\rangle_{\bx}$,
 $\triangle_{\bx} [s_{\tB}]$ and $\Psi_{\bx} [s_{\tB}]$ using (\ref{ttb1}) by
$\langle s_{\tB}\rangle$, $\triangle[s_{\tB}]$ and $\Psi[s_{\tB}]$;
that is to say, we omit the unnecessary trajectory-identifying subscript $\bx$.
If we were able
to devise a model with (i)--(iii) the same as the $N=10,000$ Kac ring, but with an ergodic dynamics,
we would have
\begin{equation}
\begin{array}{l}
\langle s_{\tB}\rangle=0.69261408,\\[0.25cm]
\triangle[s_{\tB}]= 0.49997497\times 10^{-4},
\hspace{0.7cm} \Psi[s_{\tB}]=0.70707142\times10^{-4}
\end{array}
\label{wntlt7}
\end{equation}
and it is not difficult to show that $\triangle[s_{\tB}]$
 and $\Psi[s_{\tB}]$ are monotonically decreasing functions of $N$.
Ergodicity leads to more thermodynamic-like behaviour, which becomes
increasingly thermodynamic-like with increasing $N$.\footnote{This latter result can be proved for any
division of $\Gamma_\tmN$ into macrostates with the measure given by a combinatorial quantity
like (\ref{kac1}).} This behaviour is also typical, since it occurs f.a.a. $\mcL_{\bx}\in \Gamma_\tmN$.

Of course, the results (\ref{wntlt7}) are not simply dependent on the putative ergodic dynamics of the system,
but also on the way that the macrostates have been defined. If the time average of $s_{\tB}(\bx)$
is to be close to $(s_{\tB})_\mx$ and if the fluctuations in $s_{\tB}(\bx)$ are to be small then most of $\Gamma_\tmN$
must lie in macrostates with $s_\tB(\mu)$ close to $(s_\tB)_\mx$.
 In the case of the Kac ring, with $N=10,000$,
 99.98\% of $\Gamma_{10000}$ lies in macrostates with $s_\tB(\mu)>0.999 (s_\tB)_\mx$.
However, of course, the Kac ring, although not ergodic, gives every appearance, at
least in the instances investigated (see Fig.\ \ref{fig1}), of behaving in a thermodynamic-like
manner. Although ergodicity, with a suitable macrostate structure, is {\em sufficient} for thermodynamic-like
behaviour, it is clearly not {\em necessary}.

\begin{figure}[t]
\begin{center}
\includegraphics[width=100mm,angle=0]{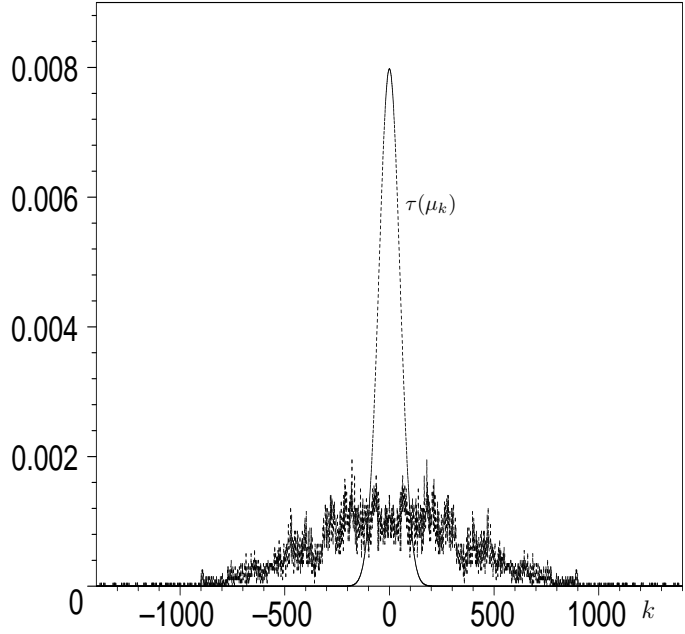}
\end{center}
\caption{Plot of $\ttime_\alpha(\mu_k)$ for the cycle $\alpha$ of
the Kac ring shown in Fig.\ \ref{fig1} together with the corresponding
curve of $\tau(\mu_k)$ for a system with ergodic dynamics.}\label{fig2}
\end{figure}

Consider the case where $\Gamma_{\tmN}$ can be ergodically decomposed; meaning that
\begin{equation}
\Gamma_{\tmN}=\displaystyle{\bigcup_{\{\alpha\}}}\, \Gamma_{\alpha},\label{ttb2}
\end{equation}
where
$\Gamma_\alpha$ is invariant and indecomposable under $\{\phi_t\}$.
Then the time spent in $\gamma\subset\Gamma_\tmN$ is
\begin{equation}
\ttime_\alpha(\gamma)=\frac{\curm(\gamma\cap\Gamma_\alpha)}{\curm(\Gamma_\alpha)},\hspace{1cm}
\mbox{f.a.a.}\hspace{0.5cm}\mcL_{\bx}\in\Gamma_\alpha,
\label{ttb3}
\end{equation}
and we, henceforth, identify
 a time average along a trajectory in $\Gamma_\alpha$ using $\langle\cdots\rangle_\alpha$
 with $\alpha$ also replacing the subscript $\bx$ in (\ref{wntlt1}) and (\ref{wntlt2}).
In the case of the Kac ring
 the index $\alpha$ labels the cycles which form the ergodic decomposition of $\Gamma_\tmN$.
 For a particular cycle, like that shown in Fig. \ref{fig1}, $\ttime_\alpha(\mu)$ is
 obtained simply by counting the number of times the phase point visits each macrostate
in a complete cycle. The data are then used to compute the results given in (\ref{wntlt5}).
A plot of $\ttime_\alpha(\mu_k)$ against the microstate index $k$ is shown in Fig.\ \ref{fig2}
with a comparison made with $\tau(\mu_k)$, the corresponding curve for a system
with ergodic dynamics. This gives a graphic illustration of the suggestion that
ergodic systems, with a suitable choice of macrostates, are likely to
be more thermodynamic-like in their behaviour than non-ergodic systems.
It might also be speculated that $\varepsilon$--ergodic systems
\citep{vran} show increasingly thermodynamic-like behaviour with decreasing
$\varepsilon$.

Another advantage of an ergodic system is that, f.a.a. $\mcL_{\bx}\in\Gamma_\tmN$,
the level of thermodynamic-like behaviour will
be the same. This contrasts with an ergodic decomposition
characterized by (\ref{ttb2}), where it is possible for differing levels
of thermodynamic-like behaviour to be exhibited within different members of the
decomposition. To be precise, take small positive $\varepsilon_\triangle$
and $\varepsilon_\Psi$ and regard behaviour along a trajectory as
thermodynamic-like if and only if both   $\triangle_\alpha[s_{\tB}]<\varepsilon_\triangle$
and  $\Psi_\alpha[s_{\tB}]<\varepsilon_\Psi$.
Let $\Gamma^{(\tT)}_\tmN$ be the union of all $\Gamma_\alpha$ in which
the behaviour is thermodynamic-like
with $\Gamma^{(\tA)}_\tmN= \Gamma_\tmN\backslash\Gamma^{(\tT)}_\tmN$.

In the discussion of the Boltzmann approach we have so far avoided any reference to probabilities.
To complete the discussion in this section and to relate our arguments to the Gibbs approach we shall need
\citep[see][]{lavis5} to introduce two sets of probabilities for a system with the ergodic decomposition
(\ref{ttb2}). The first of these is $\prbs_\alpha\cequals\prb(\bx\in\Lambda_\alpha)$, $\forall$
$\Gamma_\alpha$. Then thermodynamic-like behaviour will be typical for the system if
\begin{equation}\sum_{\Gamma_\alpha\subset\Gamma^{(\tA)}} \prbs_\alpha\ll 1. \label{ttb4}
\end{equation}
Thus we have two levels of degree, the first represented by the choices of
$\varepsilon_\triangle$ and $\varepsilon_\Psi$ and the second concerning the extent to which (\ref{ttb4})
is satisfied.
\section{Reconciling Gibbs and Boltzmann}\label{rgb}
As we saw in Sect.\ \ref{giben}, the Gibbs approach depends on defining a probability density function
$\rho(\bx)$ on $\Gamma_\tmN$, for a system `at equilibrium'. Thus we must address more directly the question
of probabilities. Assuming the ergodic decomposition (\ref{ttb2}), we take the time-average
 definition of \citet{vonP2}, for which
\begin{equation}
\prb(\bx\in\gamma|\bx\in\Gamma_\alpha)\cequals\ttime_\alpha(\gamma),\hspace{0.7cm}
\mbox{$\forall$ $\gamma\subset\Gamma_\tmN$,}
\label{rgb1}
\end{equation}
where $\ttime_\alpha(\gamma)$ is given by (\ref{ttb3}). Thus the probability density function for
$\Gamma_\alpha$ is
\begin{equation}
\rho_\alpha(\bx)=\left\{\begin{array}{ll}
1/\curm(\Gamma_\alpha), &\mbox{$\bx\in\Gamma_\alpha,$}\\[0.25cm]
0,&\mbox{otherwise,}
\end{array}\right.
\label{rgb2}
\end{equation}
from which we have
\begin{equation}
\rho(\bx)=\stackrel{\tE\tD}{\overline{\rho_\alpha(\bx)}}\,\,\cequals\sum_{\{\alpha\}} \rho_\alpha(\bx) \prbs_\alpha.
\label{rgb3}
\end{equation}
If we assume that all points of $\Gamma_\tmN$ are equally likely, then on Bayesian/Laplacean grounds, and
consonant with the approach of \citet{bric1}, we should choose
\begin{equation}
\prbs_\alpha=\curm(\Gamma_\alpha)/\curm(\Gamma_\tmN),
\label{assum1}
\end{equation}
giving, from (\ref{rgb3}), $\rho(\bx)=1/\curm(\Gamma_\tmN)$, which is the microcanonical distribution
and for which, from (\ref{giben}),
\begin{equation}
s_\tG\cequals S_\tG/(Nk_\tB)=\ln[\curm(\Gamma_\tmN)]=s_\tB(\Gamma_\tmN).
\label{rgb4}
\end{equation}
It also follows, from (\ref{bolen}), (\ref{wntlt3}), (\ref{ttb1}) and (\ref{assum1}) that
\begin{equation}
\stackrel{\tE\tD}{\overline{\langle s_\tB\rangle_\alpha}}=\langle s_\tB\rangle.
\label{rgb5}
\end{equation}
\citet{lavis5} has proposed a general scheme for relating a phase function $f$, defined on $\bx\in\Gamma_\tmN$
to a macro-function $\eusF$ defined on the macrostates $\{\mu\}$ and then to a thermodynamic function $F$.
The first step is to course grain $f(\bx)$ over the macrostates to produce $\eusF(\mu)$.\footnote{It
is argued that $\eusF$ is a good approximation to $f$ for the phase functions relevant to thermodynamics
since their variation is small over the points in a macrostate.} The second step is to define
the thermodynamic variable $F$ along the trajectory $\mcL_{\bx}$ as $\langle\eusF\rangle_{\bx}$.
In the case of the Boltzmann entropy, which is both a phase function and a macro-function
the first step in this procedure is unnecessary since it already, by definition, course grained
over the macrostates. Then we proceed to identify the dimensionless thermodynamic
 entropy per microsystem with $\langle s_\tB\rangle_{\bx}$.
In the case of a system with an ergodic decomposition this definition would yield a
different thermodynamic entropy $s_\alpha$ for each member of the decomposition, with, from (\ref{wntlt1}),
\begin{equation}
s_\alpha\cequals \langle s_\tB\rangle_\alpha=(s_\tB)_\mx-\triangle_\alpha[s_\tB].
\label{rgb6}
\end{equation}
In the case where the behavior is thermodynamic-like in $\Gamma_\alpha$,
$s_\alpha$ differs from $(s_\tB)_\mx$ by at most some small $\varepsilon_\triangle$
and, if (\ref{ttb4}) holds, this will be the case for measurements along most
trajectories. In the case of the Kac ring
with $N=10,000$ and the trajectory investigated for Figs.\ \ref{fig1} and \ref{fig2} the actual
difference is given in (\ref{wntlt5}), a value which is likely to decrease with increasing $N$.

It is often said that in ``equilibrium [the Gibbs entropy] agrees with Boltzmann and Clausius entropies
(up to terms that are negligible when the number of particles is large)
and everything is fine'' \citep[188]{bric}.
Interpreted within the present context this means that the good approximation
$s_\tB(\mu_\mx)$, for the entropy per microsystem of a system for which thermodynamic-like
behaviour is typical, can be replace by $s_\tG=s_\tB(\Gamma_\tmN)$. The advantage of this
substitution is obvious, since the first expression is dependent on the division into
macrostates and second is not. However a little care is needed in justifying this substitution.
It is not valid because, as asserted in the quote from \citeauthor{bric} on page \pageref{briclab},
$\mu_\mx$ occupies an increasing proportion of $\Gamma_\tmN$ as $N$ increases. Indeed, we have shown
for the Kac ring the reverse is the case. That proportion becomes vanishingly small as $N$
increases. However, the required substitution can still be made, since for that model
\begin{equation}
\frac{s_\tB(\mu_\mx)}{s_\tB(\Gamma_\tmN)}\simeq 1-\frac{\ln(N)}{2N\ln(2)},
\hspace{1cm}\mbox{as $N\rightarrow\infty$.}
\label{rgb7}
\end{equation}
Although it may seem that the incorrect intuition on the part of \citeauthor{bric} et al.
concerning the growth in the relative size
of the largest macrostate, leading as it does to the correct conclusion with respect to entropy,
is easily modified and of no importance, we have shown in Sect.\ \ref{hdwde} that it has
profound consequences for the attempt to define equilibrium in the Boltzmann approach.

It should be emphasized that the Gibbs entropy (\ref{rgb4}) is no longer taken as that
of some (we would argue) non-existent equilibrium state, but as an approximation to
the true thermodynamic entropy which is the time-average over macrostates of the
Boltzmann entropy. The use of a time-independent probability density function for the
Gibbs entropy is not because the system is at equilibrium but because the underlying
dynamics is autonomous.\footnote{A non-autonomous dynamic system will not yield a
time-independent solution to Liouville's equation.} The thermodynamic entropy approximated
by the Gibbs entropy (\ref{rgb4}) remains constant if the phase space remains unchanged
but changes discontinuously if a change in external constraints leads to a change
in $\Gamma_\tmN$. An example of this, for a perfect gas in a box when a partition
is removed, is considered by \citet{lavis5} who shows that the Boltzmann entropy follows
closely the step change in the Gibbs entropy.
\section{Conclusions}\label{conc}
In our programme for reconciling the Boltzmann and Gibbs approaches to statistical
mechanics we have made use both of ergodicity and ergodic decomposition and
 there is deep (and justified) suspicion of the use of ergodic
arguments, particularly
among philosophers of physics. \citet[75]{ear&red} argue  ``that ergodic
theory in its traditional form is unlikely to play more than a cameo role in
whatever the final explanation of the success of equilibrium
statistical mechanics turns out to be''.
In its `traditional form' the ergodic argument goes something
like this: (a) Measurement processes on thermodynamic systems take a long time compared to
the time for microscopic processes in the system and thus can be effectively regarded as
infinite time averages. (b) In an ergodic system the infinite time average can be shown, for
all but a set of measure zero, to be equal to the macrostate average with respect to an
invariant normalized measure
which is unique.\footnote{In the sense that it is the only invariant normalized
measure absolutely continuous
with respect to the Lebesque measure.}
The traditional objections to this argument are also well known: (i) Measurements
may be regarded as time averages, but they are not {\em infinite} time averages.
If they were one could not, by measurement, investigate a system not in equilibrium.
In fact, traditional ergodic theory does not distinguish between systems in equilibrium
and not in equilibrium. (ii) Ergodic results are all to within sets of measure zero
and one cannot equate such sets with events with zero probability of occurrence.
(iii) Rather few systems have been shown to be ergodic. So one must look for a reason for
the success of equilibrium statistical mechanics for non-ergodic systems and when it is
found it will make the ergodicity of ergodic systems irrelevant as well.

Our use of ergodicity differs substantially from that described above and it thus escapes
wholly or partly the strictures applied to it. In respect of the question of
equilibrium/non-equilibrium we argue that the reason this does not arise in ergodic arguments
is that equilibrium does not exist. The phase point of the system, in its passage along
a trajectory, passes through common (high entropy) and uncommon (low entropy) macrostates {\em and that is all}.
So we cannot be charged with
`blurring out' the period when the system was not in equilibrium. The charge against
ergodic arguments related to sets of measure zero is applicable only if one wants
to argue that the procedure always works; that is that non-thermodynamic-like behaviour
never occurs. But we have, in this respect taken a Boltzmann view. We need thermodynamic-like
behaviour to be typical and we have proposed conditions for this to be the case.
But we admit the possibility of atypical behaviour occurring with small but
not-vanishing probability.
While the class of systems admitting a finite or denumerable ergodic decomposition is likely
to be much larger than that of the purely ergodic systems,
 there remains the difficult question of determining general conditions
under which the temporal behaviour along a trajectory, measured in terms of visiting-times
in macrostates, approximates, in most members of the ergodic decomposition, to
thermodynamic-like behaviour.

\end{document}